\newif\iffigures
	\figurestrue  

\documentstyle[times,psfig]{l-aa}
\newlength{\colwidth}
\begin{document}

\thesaurus{02.01.2;02.02.1;11.10.1;11.17.3}

\title{The symbiotic system in quasars: black hole, accretion disk and jet}

\author{Alina-C\u at\u alina Donea\inst{1,2}
\and Peter L. Biermann\inst{1}}

\institute{ Max-Planck-Institut f\"ur Radioastronomie, Auf dem H\"ugel 69, 
D-53121
Bonn, Germany
\and Astronomical Institute of the Romanian Academy, Cu\c titul de Argint 
5, RO-75212, Bucharest, Romania}
\offprints{Alina Donea(adonea@mpifr-bonn.mpg.de)}

\date{received date; accepted date}
\maketitle

\begin{abstract}
  The UV continuum spectrum of quasars and AGN is assumed to originate
from an accreting disk surrounding a massive rotating black hole. We
discuss the structure and emission spectra of a disk which drives a
powerful jet. Due to the large efficiency of extracting energy from
the accreting matter in the inner part of the disk close to the
massive object, all the energetic conditions for the formation of jets
are fulfilled.  The total energy going up into the jet depends
strongly on the Kerr black hole parameters, on the disk features and
on the mass flow and thickness of the jet. The shape of UV spectra of
the AGN can be explained by a sub-Eddington accretion disk which
drives a jet in the innermost parts.


\end{abstract}

\section{Introductory}

  The general concept of an AGN (active galactic nucleus) contains
 three important ingredients. The first constituent which binds the
 space-time geometry and dictates how the matter has to move in its
 vicinity is a black hole. The great luminosities observed depend on
 the masses of the black holes $M\in[10^8 M_{\odot},10^{10}
 M_{\odot}]$. The second one is the disk of gas. Matter flows towards
 the massive object following nearly circular orbits. The disk-like
 accretion onto a black hole is the most plausible explanation for the
 strong emission in the ultraviolet (UV), the so-called "big blue
 bump" (Shields 1978) and for the X-rays in AGN. The third component
 of an AGN is the jet.  The notion that the jets and the accretion
 disks are symbiotically related originated since the first radio and
 optical observations revealed the parsec-scale of the AGNs. A
 geometrically thin disk with a mass accretion rate lower than the
 limit imposed by the Eddington accretion rate can drive outflows in
 the dense regions in the vicinity of a black hole. The flow can be
 ejected from the regions of the disk where the forces of the
 radiation pressure, gas pressure and gravitation are
 comparable. Relativistic jets can extract high amounts of energy from
 the central object (Falcke and Biermann 1995). Pudritz (1986)
 proposed that an MHD jet could extract the angular momentum of the
 accretion disk, so that the jet can control the accretion process in
 the vicinity of the last marginally stable orbit. The jet-disk system
 is governed by the conservation laws of mass, angular momentum and
 energy.

\section {The inner boundary conditions}

  We assume that the spacetime geometry where the disk develops is
described by the Kerr metric. The mass of the black hole is M and its
specific angular momentum is $a$.  The geometry of the accretion disk
is described by the radius R and its thickness H.  The disk is
geometrically thin ($ H/R \ll 1$) and we work with the basic
assumption that the accretion is quasisteady (Novikov and Thorne 1973,
hereafter NT73) with a constant rate of mass accretion $\dot{M}$. The
disk consists of fully ionized hydrogen and the effects of magnetic
fields on the disk structure are neglected in our calculations. We
consider the case that the disk is not self-gravitating and its
rotation law is keplerian.

%
\begin{figure}
\iffigures
\centerline{\psfig{figure=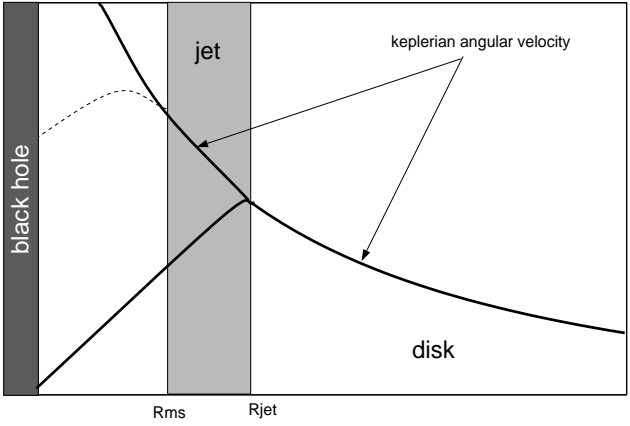,height=6.0cm,width=9.0cm}}
\else
\picplace{6.0cm}
\fi
\caption[]{\label{fig_one}Angular velocity via the radius of the disk. 
The shadowed zone indicates the innermost region of the disk where the 
jet is created. The thin-dashed line describes the behavior of the 
infalling matter if the disk can not drive the jet at its boundaries. }
\end{figure}

 We assume that the jet starts between $R_{ms}$, the last, marginally
stable orbit in the absence of a jet and the radius $R_{jet}$ with
$R_{jet}>R_{ms}$, extracting mass, energy and angular momentum from
the disk.  The presence of the jet will modify both the behaviour of
infalling matter across the radius $R_{jet}$ and the structure of the
keplerian disk.  The jets are placed in the potential well of a Kerr
black hole. The rotating central object supplies, through its
gravitational potential well, the energy necessary for driving and
maintaining a stable jet in the AGNs. We consider the case that the
gravitational potential energy released between the $R_{ms}$ and the
outer radius of the jet $R_{jet}$ goes into the jet.

On the other hand, the geometry and the total energy of the jet are
connected with the physical properties of the interior regions of the
disk. Using the above mentioned premises we have to deal with a
jet-disk symbiosis (Falcke and Biermann 1995): a ``standard disk''
(Shakura and Sunyaev 1973, hereafter SS73) and a powerful jet
(Blandford and K\"{o}nigl 1979). Both the disk and the jet structure
are governed by the mass and the spin of the black hole.

The angular velocity of the accreting matter at the innermost
boundaries will sharply decrease from $\Omega(R_{jet})$ to the angular
velocity of the black hole, if we consider that a jet exists. The
angular velocity of the black hole is unknown because the surface of
infinite redshift (the horizon) hides the image of the massive black
hole. Therefore, the massive rotating object has no solid surface on
which the accreting matter can follow. The radius of the horizon is
considered to be the surface of the black hole.
We believe that $\Omega(R)$ will drop rapidly compared with the simple
case of a disk and no jet (see Fig.1). The problem of the inner
boundary conditions for the disk structure calculations is solved by
using an approximate condition: $\frac{d\Omega(R)}{dR} = 0$ at $R =
R_{jet}$. We assume that viscous stresses do not exist across the new
boundary of disk.

In a first exploration of the physical consequences of the jet-disk
symbiosis we use $R_{jet}$ as our key parameter; it is obvious that a
full treatment would have to model the detailed
magnetohydrodynamics. However, whatever the details of any model,
energy and mass conservation have to hold, and so we parametrize a
range of possible models with $R_{jet}$, the radius where the
transition from disk to jet occurs.

\section { The connection between jet and disk}

 The unification scheme for all quasars and AGN suggests that one can
not treat the disk and jet as two separate objects. The concept of a
jet-disk symbiosis has been introduced by Falcke and Biermann (1995).
In order to describe the reciprocal influences between the three
principal objects : the black hole, the disk and the jet, we shall use
a standard set of parameters (Falcke et al. 1995a), defining:

\begin{equation} 
Q_{accr} = \dot{M} c^2,
\end{equation}
 
\begin{equation} 
q_m = \frac{\dot{M}_{jet}}{\dot{M}},
\end{equation}
  
\begin{equation} 
q_{j} = \frac{Q_{jet}}{\dot{M} c^2},
\end{equation}

\begin{equation} 
 q_{l} = \frac{L_{disk}^{jet}}{\dot{M} c^2}.
\end{equation}

Here $\dot{M}$ is the rate of mass accretion rate in the disk. The
mass flow rate into the jet is $\dot M_{jet}$. We assume that a jet
exist on both sides of the disk, and use this symmetry in our
quantitative modelling. $ Q_{jet}$ is the total power of the jet --
including the rest energy of the expelled matter -- and is expressed
as:

\begin{equation}
Q_{jet} = L_{disk} - L_{disk}^{jet}.
\end{equation}

$L_{disk}^{jet} $ is the total luminosity of a disk modified by the
presence of the jet and $L_{disk} $ is the total luminosity of the
disk if there are no physical conditions to drive the jet. A large
fraction of the total power of the jet is in magnetic fields and
relativistic particles.

The dimensionless parameters $q_{j}, q_{l}$ and $q_{m}$ have to be
less than 1, because the jet can not consume more matter and energy
than is provided by the accretion disk.
pushed out into the jet will be swallowedby the black hole.
The structure of the disk is strongly connected with a new
characteristic length scale $r_{j}$ defined as follows:

\begin{equation}R_{jet}=R_{ms} (1+r_{j}).\end{equation}

The parameter $r_{j}$ is dimensionless and the null values for the
$q_{m}$ and $r_{j}$ require the simple system with two components: the
black hole and the disk.

\section{ Properties of the Kerr disk driving jet }

 For radii less than $30 R_{g}$, where $R_{g}=GM/c^{2}$ is the
gravitational radius we need to introduce the relativistic corrections
(NT73) in order to describe the accretion disk around the Kerr black
hole and the relativistic collimated jet.

We follow the methods of standard accretion disk theory (SS73),
splitting the calculation of the disk structure into four parts: the
analysis of mass, energy and angular momentum conservation for the
disk-jet system, the structure of the disk and the propagation of
radiation through the disk's z--layering. The z coordinate is parallel
to the symmetry axis of the entire jet-disk system and perpendicular
to the disk. The geometry of the disk is evaluated using the
dimensionless parameters : $r_{\ast} \equiv R/R_{g}$ and $r_{jet \ast}
\equiv R_{jet}/R_{g}$. The specific angular momentum of the black hole
is represented by $a_{\ast}=a/M$.

We use the general relativistic correction factors denoted in Page and
Thorne (1974), hereafter PT74, and NT73 as ${\cal A}$, ${\cal B}$,
${\cal C}$, ${\cal D}$, ${\cal E}$, ${\cal F}$, ${\cal G}$, ${\cal J}$
in order to compute the quantities which describe the disk-jet
symbiosis. These parameters have been used for simplicity of splitting
formula into Newtonian part plus the relativistic correction. They are
functions of $r_{\ast}$ and the parameter $a_{\ast}$. For $R\gg
R_{ms}$ all factors mentioned tend to unity.

The radial structure of the disk is obtained by using the laws of
conservation of rest mass, angular momentum and energy.

\subsection{Rest-mass conservation}

The rest-mass flowing inward through a cylinder of radius R in the
time interval $\Delta t$, taking into account the flow along the jet
is $ \dot{M}\Delta t$, where $\dot {M}$ is:

\begin{equation}
\dot{M} = -2 \pi R \ \Sigma \ v^{\hat R} \  {\cal D}^{\frac{1}{2}} + 
\dot{M}_{jet}.
\end{equation}

The radial function $\Sigma$ is the surface density in the disk and
$v^{\hat R}$ is the mass averaged infall velocity.

\subsection {Angular--momentum conservation}

The disk-like accretion is possible only if the angular momentum can
 be transported through the disk by some effective viscosity.  The
 $\alpha$ viscosity law follows the prescriptions of SS73.  The
 viscous torque is proportional to the pressure in the disk which
 depends on the jet and black hole parameters. We used a general
 constant value of 0.1 for the $\alpha$ parameter.

 The additional loss of angular momentum into the jet modifies the 
conservation law of angular momentum in the disk as follows :

\begin{displaymath}
\bigg[-\dot{M}(1-q_{m})\  \tilde L + 2 \pi \ R^2 {\cal B} 
{\cal C}^{-1/2} {\cal D W}\bigg]_{,R}+\ 4 \pi\  R\  \tilde L\  F +
\end{displaymath}

\begin{equation}
+ \ 2 \pi \ R \ \dot{M}_{jet} \ \tilde {\tilde L}_{jet} = 0, 
\end{equation}

 The integrated shear tensor $\cal{W}$ is similar to that defined in
NT73. $F$ is the flux of emitted energy through the upper (lower) face
of the disk and $\tilde {\tilde L}_{jet}$ is the specific angular
momentum extracted by the jet per unit surface.

The first term in the brackets is the angular momentum carried by the
remaining mass of gas towards black hole determining the spin of the
central object. The second term represents the angular momentum
transported by the torques in the disk. The third and the fourth terms
contain information about the angular momentum carried away by the
radiation and, respectively, by the jet.

\subsection {Conservation of energy}

The gravitational potential energy is released in the disk through
viscous dissipation of shear generated by differential rotation of the
disk. The energy flux in the vertical direction is:

\begin{equation}
F=\frac {3}{4} \bigg(\frac {1}{M r_{\ast}^{3/2}} \bigg) 
\frac{{\cal D W}}{{\cal C}}
\end{equation}

Using the conservation laws for energy and angular momentum we get a
differential equation for the integrated stress ${\cal W}$ with an
additional term for the angular momentum extraction by the jet:

\begin{displaymath}
\bigg[-\dot{M}(1-q_{m})\  \tilde L + 2 \pi \ R^2 {\cal B} 
{\cal C}^{-1/2} {\cal D W}\bigg]_{,R}+
\end{displaymath}

\begin{equation}
+\ 2 \pi\  \frac {3}{2} \bigg(\frac {M}{R}\bigg)^{1/2} \tilde L 
\frac{{\cal D}}{{\cal C}}{\cal W} +\ 2 \pi \ R \ \dot{M}_{jet}
 \ \tilde {\tilde L}_{jet} = 0, 
\end{equation}

The solution of the equation is straightforward. In order to solve
more easily the differential equation (10) and to keep a similar form
of the final result as that found by NT73, we define ${\cal L^{'}}$
and ${\cal Q^{'}}$, two functions of $r_{\star}$. Only, when we
consider no-jet models we can get a form of these parameters equal to
those of NT73. Due to the consideration of the jet in our model, the
equation (10) includes the angular momentum extracted by the jet. The
parameters: ${\cal L^{'}}$ and ${\cal Q^{'}}$, depend on the presence
of the jet, also. Hence,

\begin{equation}
{\cal L^{'}} \equiv \frac{\tilde L(r_{\ast})-\tilde 
L(r_{jet \ast})}{M r_{\ast}^{1/2}},
\end{equation}

\begin{equation}
{\cal Q^{'}}   \equiv   (1-q_{m}){\cal L^{'}}-\frac{3}{2 r_{\ast}^{1/2}} 
{\cal J}\tilde I
\end{equation}
where $\tilde I$ is:

\begin{equation}
\tilde I  \equiv  \int_{r_{jet\ast}}^{r_{\ast}} \bigg [(1-q_{m})\frac
 {{\cal F}}{{\cal B C}} \frac{{\cal L^{'}}}{{\cal J}} \frac{1}{\tilde 
r_{\ast}^{3/2}} + \frac{2}{3} \frac{q_{m}}{M}
\frac{\tilde {\tilde L}_{jet}}{\cal J} \bigg] d\tilde r_{\ast}
\end{equation}

$\tilde {L}(r_{jet})$ is the angular momentum per unit mass for the
circular orbit at $R=R_{jet}$.  The case of the disk without the jet
is obtained if $r_{j}{\to 0}$, $q_{m}{\to 0}$ and if the inner
boundary is fixed by the limit $r_{jet\ast}{\to r_{ms\ast}}$. The
dimensionless parameter for the innermost stable orbit is $r_{ms\ast}
\equiv R_{ms}/R_{g}$.

The constant of integration is found from the condition that no
viscous stresses act across the boundary $R=R_{jet}$. The amount of
the energy dissipated per unit area in unit time is computed using the
new ${\cal L^{'}}$ and ${\cal Q^{'}}$ parameters defined above and has
a similar form comparing to that found by NT73 :

\begin{equation}
F=\frac{3 G}{8 \pi}\frac{\dot{M}}{M^2} \frac{1}{r_{\ast}^3}\frac{{\cal Q^{'}}}{{\cal B}{\cal C}^{1/2}},
\end{equation}
with the new ${\cal Q^{'}}$ and ${\cal L^{'}}$ parameters defined above.

In Fig. 2 we show the energy dissipation released into the disks 
surrounding rotational or stationary black holes. The thick lines 
correspond to disks placed in a Kerr geometry.

\begin{figure}
\iffigures
\centerline{\psfig{figure=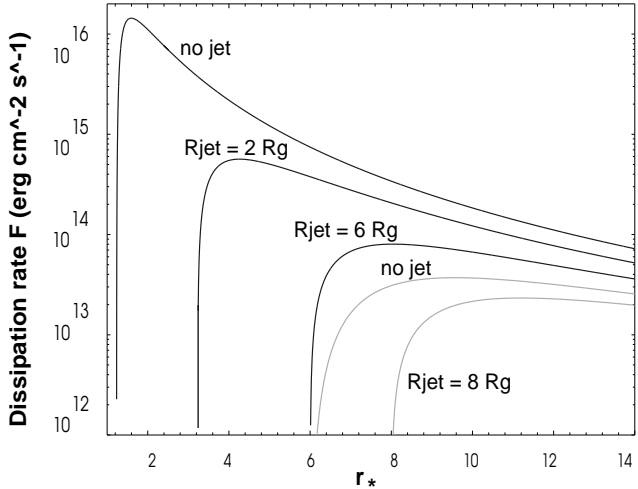,height=7.0cm,width=9.0cm}}
\else
\picplace{6.0cm}
\fi
\caption[]{\label{fig_two}The energy dissipation rate F against the
dimensionless parameter $r_{\ast}$. The thick lines correspond to 
the case of a maximally rotating black hole and the thin lines
 correspond to the Schwarzschild black hole. The mass of black 
hole is $M=10^8 M_{\odot}$ and $R_{jet}$ is the outer radius of 
the jet defined in text.}
\end{figure}

The total luminosity from the two sided-disk which gives birth to the
jet is given by integration over the disk surface:

\begin{equation}
L_{disk}^{jet}=4 \pi \int_{R_{jet}}^{R_{out}} F(R) R dR
\end{equation}

 If we keep a constant mass $M$ and a constant mass accretion rate
$\dot{M}$ then it is obvious that for models with the same rotation
parameter, the liberated fluxes are smaller for systems driving jets
compared with those that do not (see Fig. 2). We plot the energy
fluxes released by the Kerr disk with a jet for two different outer
radii of the jet: $R_{jet}=2R_{g}$ and $R_{jet}=6R_{g}$.  The bigger
the base of the jet, the larger the amount of energy going into jet.

If one asks for a geometrical disk with the innermost Keplerian orbit
at $6 R_g$, we can choose the system consisting of: maximal Kerr black
hole, a jet starting between $1.23 R_g$ and $6 R_g$ and the outer disk
(see Fig. 2).  One can see that this particular model replaces the
simple case of a Schwarzschild black hole and a disk without jet. The
comparison of these two models leads to the conclusion that the energy
liberation rate in the disk with the jet surrounding a Kerr black hole
is greater than the energy dissipation rate in a simple disk
surrounding a stationary black hole, when both the disks have equal
inner radii. The release of gravitational energy can be explained with
these symbiotic systems. This is the key result which allows the fits
to observed spectra shown below.

 The precipitous drops in the radiative flux, when jets are present
for different spins of black holes are straight results emerging from
our assumptions that the gravitational potential energy released
between the $R_{ms}$ and the outer radius of the jet $R_{jet}$ goes
into the jet. This sharp drop is easily understood, if one considers
the Kerr disk as reference: it arises from cutting a Kerr disk off at
$R_{jet}$. The Keplerian disk does not exist anymore inside $R_{jet}$.

\section{Spectrum of the disk}

 The disk has three distinct regions pointed out by SS73: the dominant
radiation pressure inner zone, the dominant electron scattering zone
and the gas pressure and free-free absorption external zone.

The boundary radii describing the transitions between the three
distinct regions of the relativistic disk with the jet surrounding a
maximal Kerr black hole are smaller than those found for the disk
which does not drive the jets in the same spacetime.  It means the
regions where the radiation pressure dominates are small, and extend
to moderate radii close to the footpoint of the jet.

The radiation pressure can not dominate the gas pressure in the
interior zones of the disks, if the mass accretion rates are small and
the outflowing jets have thick bases and large mass fluxes.

 The final analytical formulae for the complete structure of the
relativistic disk are similar to those derived in NT73 but with our
new ${\cal L^{'}} $ and ${\cal Q^{'}}$ parameters replaced, for all
three regions with different physical properties.  If the disk is
optically thick and absorption dominates over scattering, the local
emission from the disk can be described by a Planck blackbody function
with an effective temperature:

\begin{equation}
T_{eff}=\bigg(\frac{ F(r_{\ast})}{\sigma}\bigg)^{1/4}
\end{equation}

In Fig.3 we plot the density $\rho_{0}$, the effective temperature
$T_{eff}$ and the $z_{0}$ half-thickness for different innermost zones
of the relativistic disks. We take $\dot {M} = 0.01 M_{\odot}/yr$, the
mass of the black hole $M=10^8 M_{\odot}$ and the dimensionless
parameter of the specific maximum angular momentum $a_{\ast} =
(0,0.9981)$. The structure of the disk with no jet and that of a disk
with a jet starting between $6 R_g$ and $10.77 R_g$, surrounding a
Schwarzschild black hole is shown in Fig 3. with thin lines.

\begin{figure}
\iffigures
\centerline{\psfig{figure=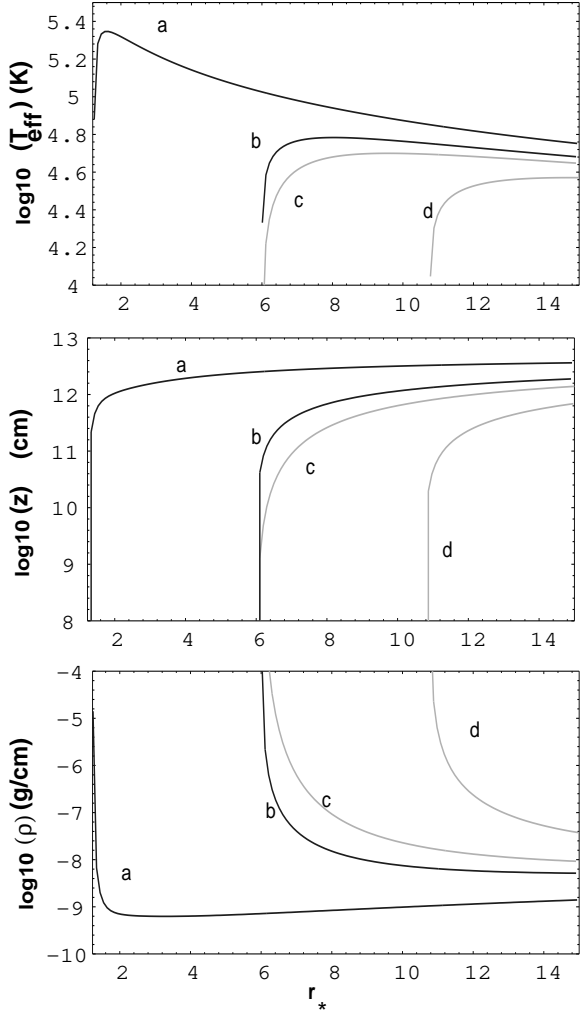,height=14.0cm,width=9.0cm}}
\else
\picplace{13.0cm}
\fi
\caption[]{\label{fig_three} The effective temperature $T_{eff}$, 
the density $\rho_{0}$ and the half-thickness $z_{0}$ of the 
inner part of the disk as functions of the dimensionless 
radius $r_{\ast}$ for the similar cases mentioned in the Fig 2. 
The darker lines show the cases of the disks surrounding a Kerr 
black hole and the grey lines show the disks in Schwarzschild 
geometry. The cases a) and c) show the properties of the disk 
driving no jets. The cases b) and d) correspond to the jets 
with equal thicknesses of $4.77Rg$. }
\end{figure}

As we expect, one observes in Figures 2 and 3, that the simple
 rotating models containing only a Kerr black hole and a disk (case a)
 are luminous and hotter than those with stationary black holes (case
 c).  The case b) corresponds to the system: a Kerr black hole, disk
 and a jet between $1.23R_g$ and $6R_g$.

By comparison of the case a) with outflow and the case b) with no
outflow, we see that the disk is denser and thinner in the second
case. We use the same rotation parameter $a_{\star}$. The thicker the
base of the jet the thinner is the new disk. The effective
temperatures could decrease by one order in magnitude. A temperature
of $10^{4.5}$K in the radiation pressure zone is high enough in order
to explain the Big UV-Bump observed in the AGNs spectra.

A jet between $1.23R_g$ and $6R_g$ leads to values of the densities
and thickness of the new structure in the disk, similar to those
computed for a simple disk in Schwarzschild geometry. These
similarities permit to replace the simple systems, with the new one
consisting of a powerful jet, a disk like and the central object.

The case d) in Fig. 3 shows the properties of a disk surrounding a
Schwarzschild black hole. The thick jet in this example has the inner
radius at $6 R_g$ and the outer radius at $10.5 R_g$.

The energies of the emitted photons from the disk are dependent on the
opacities, the temperatures and the densities in the disk. At high
frequencies the opacity in the disk is dominated by Thomson scattering
and for the inner regions of disk the comptonization will modify the
spectrum if the $y$ parameter is large enough.  In the inner regions
of the disk the Compton parameter:

\begin{equation}
y=\frac{4 k T_{e}}{m_{e} c^2} \tau_{T}^2
\end{equation}
has to be greater than 1 and depends on the electron temperature and
the Thompson optical depth $\tau_{T}=k_{T} \rho z_{0}$ where $k_{T}$
is the Thompson opacity coefficient.  The mass and energy loss into
the jet lead to a shrinking region of electron scattering dominant
opacity. A thick base of the jet can cover the whole region in which
comptonization can harden the blackbody photons emitted in a disk
which has no jet.

\begin{figure}
\iffigures
\centerline{\psfig{figure=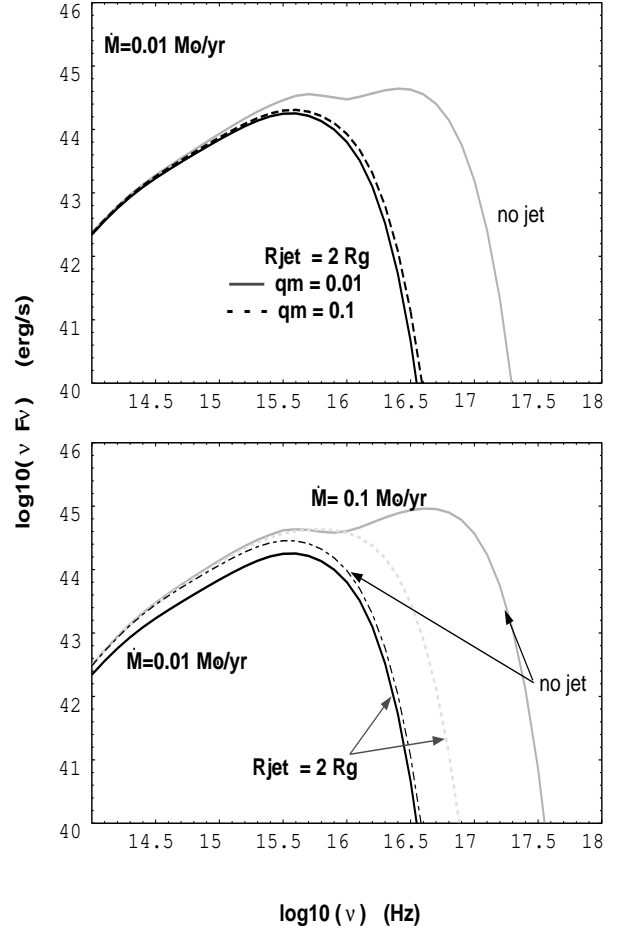,height=14.0cm,width=9.0cm}}
\else
\picplace{6.0cm}
\fi
\caption[]{\label{fig_four} Emergent spectra for 
$M=10^8 M_{\odot}$ and $\dot{M}=0.01M_{\odot}/yr$. The light 
colored lines show the computed emergent spectra when the 
disk does not drive a jet, for different mass acretion rates. 
The continuous dark lines show the computed emergent spectrum 
when the disk drives a jet with $R_{jet}=2 R_g$, and  
 $q_m=0.01$. The thin dark dotted line in the second graph 
shows a spectra emerging from a disk that does not drive a jet, 
for a small $a_{\star}=0.7.$}
\end{figure}

It is not clear if the disk is thermally unstable in the inner regions
of the relativistic disk. There are some questions which have to be
solved: does the jet make these inner regions thermally unstable or
does an initial thermally unstable disk become thermally stable due to
the presence of the mass flowing into the jet? Are there the viscous
instabilities in the relativistic disk with the jet?

\begin{figure*}
\iffigures
\centerline{\psfig{figure=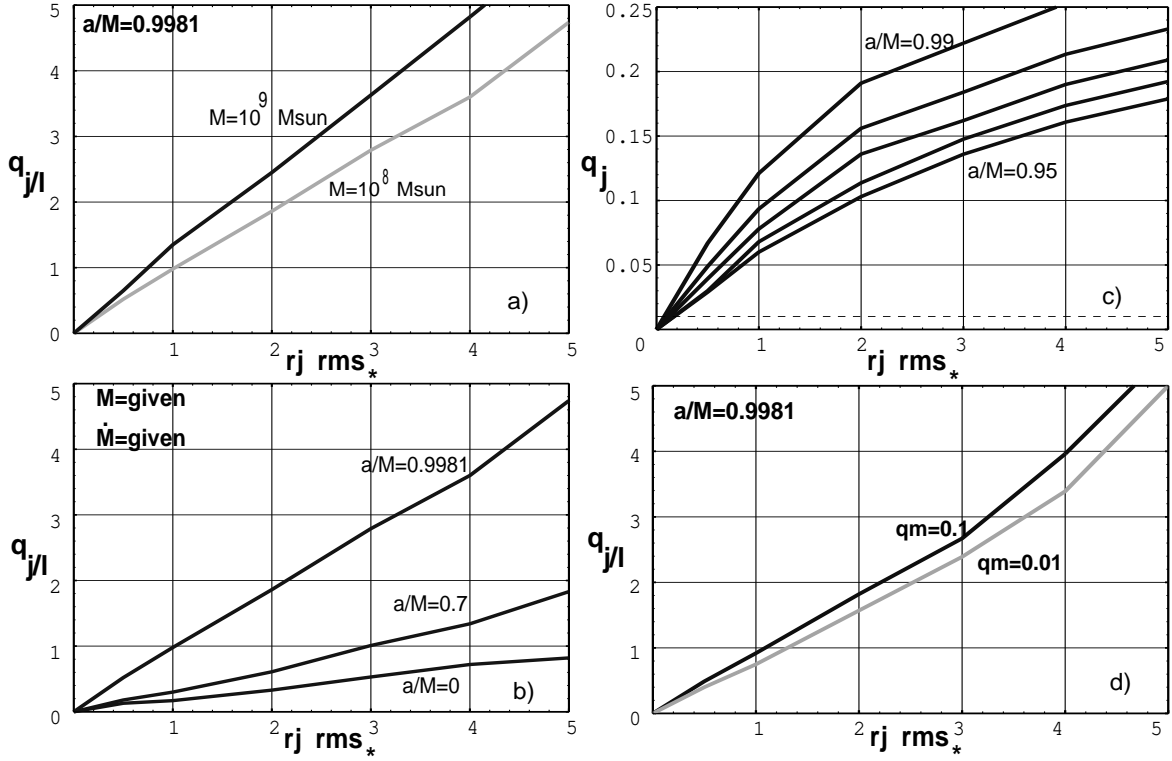,height=12.0cm,width=17.50cm}}
\else
\picplace{12.0cm}
\fi
\caption[]{\label{fig_five}  Fig 5a: $q_{j/l}$ versus
 $r_{j} r_{ms\ast}$ for different masses M of black holes. 
Fig 5b: $q_{j/l}$ versus $r_{j} r_{ms\ast}$ for different 
spins of the black hole. Fig 5c: $q_{j}$ versus $r_{j} 
r_{ms\ast}$ if the specific angular momentum of the black hole 
varies slowly  between the values $[0.95,0.99]$. Fig 5d: $q_{j/l}$ 
versus $r_{j} r_{ms\ast}$ for different mass flowing into the jet 
per unit time.}
\end{figure*}

The relativistic effects and the presence of jet modify the locally
emitted spectrum. In Fig. 4 we plot the dependence of the emitted
spectrum of the parameters $q_m$ and $r_j$. In the optical UV range
the general form of the spectra is not significantly changed when the
jets grow close to black holes. Most of the UV radiation originates
from the inner region of disk outside the $R_{jet}$. Due to our
theoretical assumption about the energy source of the jets, the form
of the spectrum coming from an "old standard" disk rotating around a
maximally rotating black hole is cut off at high frequencies, from EUV
to soft X-ray range.

 Our model containes two free parameters: $q_m$ and $r_j$. The other
two parameters $q_j$ and $q_l$ (relations (3) and (4)) concerning the
energetics of our systems, are functions of $q_m$ and $r_j$ as we will
see in the next section.

The range of variation of the first free parameter has to be between 0
and 1 (see Section 3). From the relativistic parametrized Bernoulli
equation for the jet (Falcke et al. 1995b) one may deduce, if the jet
velocity of the jet is known, the mass ratio $q_{m}$ pushed out into
the jet. We believe that a highest possible parameter $q_{m}$ can be
close to $0.1$. Therefore, the emitted spectrum depends weakly on the
parameter $q_m$ (see Fig. 4).

The flattening of the spectra in the UV range due to the interactions
between the blackbody photons and electrons in the disk is slightly
accentuated if we choose a thick base for the jet and a maximal ratio
$q_{m}=0.1$.

The second free parameter $r_j$ provides information about the
geometry of the footpoint of the jet. The range of variation of $r_j$
is limited by the straight condition of the case no-jet, where
$r_{j}=0$. From the observations we know what values should be
expected for the UV luminosities for a certain quasar. A thicker base
of the jet diminishes the integrated flux emerging from the disk. So,
one can not choose any maximum possible values for the outer radius of
the jet because the constraints introduced by the data points.

One can see from the graphs that emergent spectrum from the simple
model of a disk surrounding a rotational black hole with $a_{\star} =
0.7$ and a mass $M=10^8 M_{\odot}$ is similar with the spectrum coming
out from a Kerr disk with the jet.

We have many possibilities to explain the same UV data of the quasars
with the general model consisting of: black hole, a disk and a jet. A
higher accretion mass rate in the disk ($\dot{M}=0.1 M_{\odot}/yr$)
and a jet growing (see Fig.4b) between $1.23 R_g$ and aproximately $3
R_g$ determine a final form of the emitted spectrum identical with
that computed for the $\dot{M}=0.01 M_{\odot}/yr$ and $R_{jet}=2
R_g$. We can also reach a similar form if we take the simplest model
of a black hole with a smaller spin ($a_{\star}=0.7$) and a thin disk
(see Fig.4). We can therefore obtain good fits at different mass
accretion rate and different radii of the jet in the maximal Kerr
geometry.

\section{ The energetics of the jet}

 We consider that the flow of matter absorbed by a supermassive black
hole is the primary source for mass outflows in the inner dense part
of a disk.  Many different types of mechanisms have been proposed for
the formation of the jets. We note here the jet model driven by
thermal or radiation pressure (Lovelace et al. 1994) or the formation
of the jet due to the magnetic activity of the accretion disks
(Camenzind 1986). We do not wish to propose an explanation of the jet
formation. We show that some information about the size of the
formation region and the energy of the jet can be inferred studying
the emitted flux from the accreting disk.

We have defined in the third section $q_{j}$ as the ratio between the
total power gained by the jet and the rest energy at infinity of the
matter $\dot {M} c^2$. The Fig. 5 shows how the total energy of the
jet is connected with its geometrical parameter $r_{j}$. The greater
the specific angular momentum and the mass of black hole, the more
powerful is the jet (see Figs. 5a and 5b).

The dimensionless parameter $q_{j/l}$ defined below, gives
the ratio  between the total jet power and the 
disk luminosity:

\begin{equation}
q_{j/l}=\frac{Q_{jet}}{L_{disk}^{jet}}
\end{equation}

 The total power gained by the jet $Q_{jet}$ varies with the size of
the starting area, respectively with $r_{j}$ and depends slowly on the
dimensionless parameter $q_{m}$ which gives information about the mass
loss rate into the jet (see Fig. 5d). The disk luminosity
$L_{disk}^{jet}$ is equal to the total energy pushed into the jet
$Q_{jet}$ if the jet has a thickness given by the product $r_{j}\cdot
r_{ms\ast}=1$, where $r_{ms\ast}=1.23$ for a maximal Kerr black
hole. A large accretion rate of the disk makes the jet strong. At the
boundary zone, the structure of the old disk is destroyed. A fast
rotating starved black hole attracts the gas matter so strongly that a
big fraction $(1-q_{m})\dot M$ can not be saved by the jet from the
inevitable swallowing.

We imagine that at the formation of the escaping jet the flow occurs
through a cylinder with $R_{min}=R_{ms}$ and $R_{out}=R_{jet}$ (we
shall call that "base of the jet" ) and then begins to expand covering
the inner part of the disk.

The kinetic energy in the jet has to be great enough in order to
provide all sources of energy required by the nonthermal
processes. The total power going into the jet is limited by the
Eddington disk luminosity. The energetic processes which we assume to
occur in the jet are: particle acceleration by shocks, hadronic
interaction between energetic protons and photons from the disk
(Mannheim 1995) as well as between energetic protons and the accretion
disk (Niemeyer and Biermann, 1996). In this latter model the energetic
protons are created at the boundary of a nuclear jet and then diffuse
towards the disk to finally produce the hard X-ray spectra of the
Seyfert galaxies. The hadronic reactions require a powerful jet which
should be able to transport energy far away from the disk equatorial
zone.

We do not include in our computation the influence on the disk
spectrum coming from a possible hot corona (Haardt and Maraschi 1991)
which should be situated on the top of the disk layer. We do not know
the height of the base of jet which can be totally immersed in an
optically thin, hot corona and therefore can influence the comptonized
UV disk photons.

\section { Theory and observations}

A study of a large sample of quasars suggests that our model agrees
with data. We present here the sample of the six PG quasars examined
in the soft and hard X-ray domains by Rachen et al. (1995, hereafter
RMB95).

\begin{figure}
\iffigures
\centerline{\psfig{figure=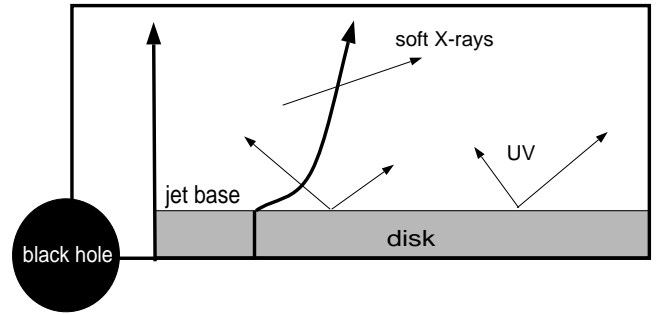,height=5.0cm}}
\else
\picplace{5.0cm}
\fi
\caption[]{\label{fig_six} The simple geometry of the symbiotic 
system: black hole, accretion disk and the jet.}
\end{figure}

We adopt Mannheim's model (Mannheim et al. 1995) to explain the
existence of the soft X-ray emission (see Fig. 6). If the base of jet
is surrounded by the UV radiation field it is expected that these
photons will heat the base of jet, and after repeated
Compton-scatterings with the hot thermal electrons in jet, they will
become hardened. The unsaturated componization will produce a
power-law extension of the UV bump spectrum. The resulting spectral
index depends sensitively on the Compton $y$ parameter which must be
close to 0.5. The thickness of the jet limits the maximum energy of
the UV photons which will hit the base of the jet.

\begin{figure*}
\iffigures
\centerline{\psfig{figure=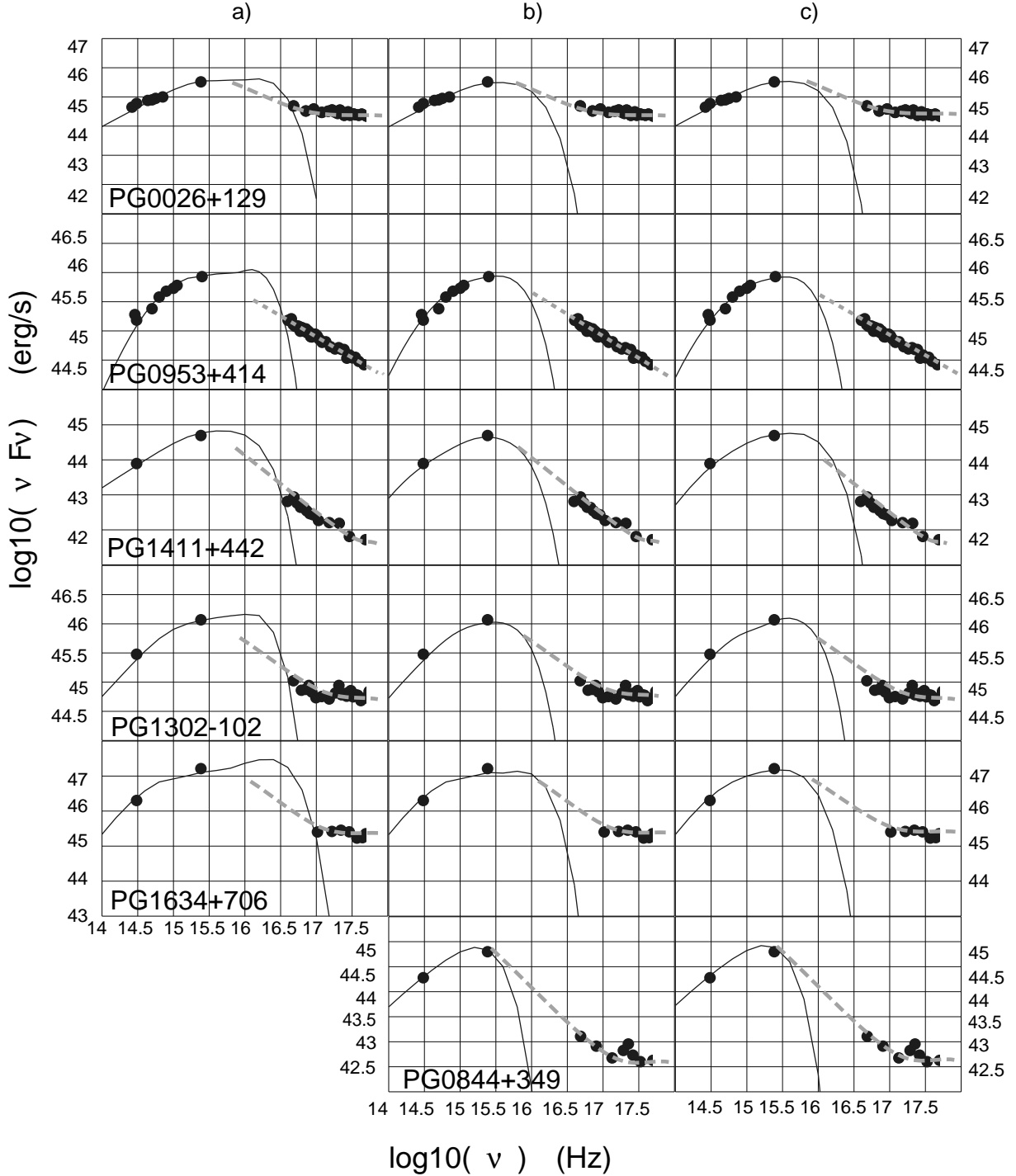,height=19.5cm,width=18.0cm}}
\else
\picplace{19.5cm}
\fi
\caption[]{\label{fig_seven}
Spectral energy distribution of the PG quasars in their source
frame. The solid lines show the theoretical spectra of the
relativistic accretion disk driving the jet in the interior
regions. The dashed curves try to fit the X-ray data taking into
account the the break energy where the soft and hard X-ray power law
intersect (see text). The a) case plots the luminosity of an accretion
disk with the jet surrounding a maximal Kerr black hole so as to have
an intersection point between the fit-curve of the spectrum of the
disk and the soft X-ray data. In the b) case we have a powerful
extreme jet. The intersection points between the solid lines and the
dashed curves are "pivot point" for all lines which try to fit the
soft X-ray data. The c) case represents the luminosities of the disk
around a slowly rotating black hole for the simple case of a disk
which does not drive a jet.  The UV continuum data are from Mannheim
et al. (1995), Malkan (1989) and Laor (1990). The soft X-ray data
points are from the RMB95. The Hubble constant adopted has the value
$H_{0}=50$ km/s/Mpc.}
\end{figure*}

 The X-ray data of the all six quasars can be fitted very well by
double power law spectra. RMB95 pointed out that there is a soft X-ray
power law spectrum with an averaged observed spectral slope :
$\alpha_{sx} = 2.1\pm0.3$. A flatter hard X-ray power law
($\alpha_{hk}\simeq 1$ for radio-weak AGN and $\alpha_{hk}\simeq 0.7$
for radio-louds) cross with the soft X-ray power law at a break energy
close to the value $1.4\pm1.1$KeV. \ Friedrich (1994) gives detailed
information about the X-ray data of the quasars, including a
statistical description of the soft and hard power law slopes.

 Our model fits well all the data in the UV range if we choose large
masses for rotating black holes, sub-Eddington accretion rates and
powerful jets with $r_{j}\cdot r_{ms \ast} \in [0.2,5.5]$.  The first
case a) in Fig 7. corresponds to the "Kerr maximum" case because we
have chosen the dimension of the escaping jet so that we wish to have
an intersection point between the fit-curve of the spectrum of the
disk and the soft X-ray data. We assume that the disk photons with
energies above the energy corresponding to the intersection point will
be comptonized in the jet producing the slope observed in the soft
X-ray domain of the majority of quasars.
 
of these 
six quasars. 
X-ray data of the 
soft and hard power law 
hard X-ray power law slope 
and $\alpha_{hk}\simeq 0.7$ for 
data situated below the break energies which 
soft and hard X-ray power law intersect (RMB95), in 
connection between our model and the Mannheim's model. Above 
energies a horizontal line will fit the hard X-ray data points.

\begin{table*}
\begin{small}
\caption{\label{tab_one} The fitting parameters of the 6 PG quasars 
for the three
cases : a) Kerr maximum, b) Maximal jet, c) No jet and slowly rotating
black holes}
\begin{tabular}{llllrrrrr}\hline

Source name & $log_{10}Q_{jet}$ & $log_{10}L_{disk}^{jet}$ & $ q_{j}$
& $q_{j/l}$ & $\dot{m}$ & $a_{\ast}$ & $r_{j}\cdot r_{\ast ms}$ &
cases
\\[.2cm]\hline

PG0953+414 & 46.59 & 46.54 & 0.23 & 1.09 & 0.25 & 0.9981 & 1.2& kerr
           maximum\\ & 46.73 & 46.30 & 0.32 & 2.67 & 0.25 & 0.9981 & 3
           & maximum jet\\ & & & & & 0.25 & 0.8 & 0 & no jet \\\hline
           PG0026+129 & 45.50 & 46.10 & 0.09 & 0.25 & 0.08 & 0.9981 &
           0.2& kerr maximum\\ & 45.92 & 45.87 & 0.24 & 1.10 & 0.08 &
           0.9981 & 1 & maximum jet\\ & & & & & 0.08 & 0.95 & 0 & no
           jet\\\hline PG1411+442 & 45.90 & 45.21 & 0.35 & 4.80 & 0.58
           & 0.9981 & 5.5& kerr maximum\\ & 45.95 & 44.89 & 0.39 &
           11.4 & 0.58 & 0.9981 & 10 & maximum jet\\ & & & & & 0.58 &
           0 & 0 & no jet\\\hline PG1302-102 & 46.88 & 46.62 & 0.27 &
           1.81 & 0.42 & 0.9981 & 2 & kerr maximum\\ & 46.96 & 46.23 &
           0.33 & 3.42 & 0.42 & 0.9981 & 4 & maximum jet\\ & & & & &
           0.42 & 0.8 & 0 & no jet\\\hline PG1634+706 & 47.80 & 47.90
           & 0.19 & 0.79 & 0.57 & 0.9981 & 0.8& kerr maximum\\ & 48.02
           & 47.60 & 0.31 & 2.64 & 0.57 & 0.9981 & 3 & maximum jet\\ &
           & & & & 0.57 & 0.6 & 0 & no jet \\\hline PG0844+349 & 45.95
           & 45.17 & 0.39 & 6.06 & 0.42 & 0.9981 & 5 & maximum jet\\ &
           & & & & & 0.4 & 0 & no jet\\\hline
\end{tabular}
\end{small}
\end{table*}

Using the unsaturated comptonization at the base of jet we could find
other possible solutions in order to explain the continuity from the
"big blue bump" to the hard X-ray domain.

The second case b) from the Fig. 7 belongs to the "maximal jet"
possibility. In this figure we plot with dashes the possible curves
which fit the X-ray data using the general averaged slopes emphasized
by RMB95 and the respective break energy values of each quasar. The
tangential line corresponding to the fit of soft X-ray data intersects
the theoretical curve of the spectrum of the disk in a "pivot"
point. This point can be the maximum of the luminosity of the
disk. According to Mannheim's model the physical conditions at the
base of the jet can influence the Compton parameter, implicitly the
slope of the fit line in the soft X-ray range.  We have a good fit of
the UV data for all 6 quasars, in concordance with the disk model and
a thicker base of the jet with $r_{j} \cdot r_{ms \ast} \in
[1,10]$. The spectra emerging from the disk are cut off at high
energies having a maximum in the UV range. The value of the frequency
corresponding with this maximum $\nu_{max}$ is correlated with the
$q_{m}$ and $r_{j}$ parameters. A thick base of the jet diminishes the
value of the frequency $\nu_{max}$.  The bigger the energy pushed into
the jet, the smaller is the maximum photon energy coming from the
disk.  We can find the powerful possible case for the jets near a Kerr
black hole with the specific angular momentum $a_{\ast}$.

The disk spectrum becomes softer when the black hole is rotating at
$a_{\ast}<0.9981$.  That means, the jets will be poorly fed.  The last
c) case represents the luminosities of the disk around a slowly
rotating black hole for the simple case of a disk which does not drive
a jet in the innermost parts. We plotted the most extreme possible
cases which can explain the UV data for quasars. This case requires
the choice of the other models in order to explain the presence of the
X-ray in quasars.

In Table 1 we present all parameters of the jets deduced for the 6 PG
quasars.  The dimensionless accretion rate $\dot{m}$ is defined
$\dot{m} \equiv \dot{M}/\dot{M}_{edd}$, where $\dot{M}_{edd}$ is the
Eddington mass accretion rate.
\begin{figure}
\iffigures
\centerline{\psfig{figure=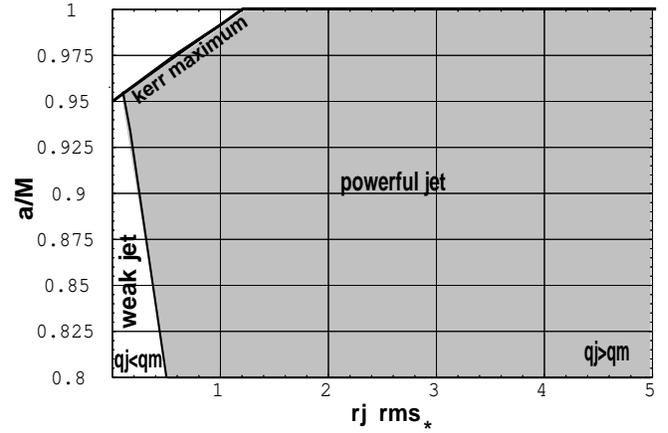,height=6.0cm,width=9.0cm}}
\else
\picplace{6.0cm}
\fi
\caption[]{\label{fig_8}
The permitted area for a powerful jet starting close to a fast
rotating black hole. Using our model we find that the minimum specific
angular momentum possible for the black hole in the PG0953+414 quasar
is dictated by the UV data. An angular momentum per unit mass
$a_{\ast}$ less than $0.8$ is not allowed if we fit the UV data
points. }
\end{figure}

The radio-weak quasar PG0844+349 with a low redshift (z<0.1) has the
disk luminosity of $1.4\cdot 10^{45}$ erg/s. We found that the jet
extracts 40\% of the total energy of the accreting matter. For a
maximal Kerr black hole we get an extreme powerful jet with a thick
base ($R_{jet} = 5R_{ms}$)
and with its total energy $Q_{jet}=9\cdot 10^{45}$ erg/s. If we assume that the disk can not drive a jet in its dense regions and the black hole has a specific angular momentum wich correspond to the value $a_{\ast}=0.4$ and a mass of $6\cdot 10^8 M_{\odot}$ the UV spectrum of this quasar can be explained too.

We work with sub-Eddington accretion rates in order to be in
 concordance with the starting assumption of a geometrically thin
 keplerian disk.  A simple black body emission disk model which is
 rotating around a Schwarzschild black hole is not sufficient to
 explain the large UV fluxes for the quasars. In such a picture a
 large number of quasars require super-Eddington accretion rates.

In Fig. 8 we plot the permitted area for a strong jet in the PG
0953+414 quasar. The thick oblique line corresponds to the "kerr
maximum" case for different specific angular momentum of black holes
and different thickness of the base of jet. The shadowed zone covers
the solution parameters which determine a good fitting of the UV data
of the PG quasars if we take into account the comptonization of the
radiation on the hot thermal electrons at the base of the jet.

  We believe that only for fast rotating black holes with
$a_{\ast}>0.9$ and $r_{j}\cdot r_{ms \ast} > 1$ we may obtain the
proper conditions at the base of jet to get unsaturated comptonized
photons.  For slowly rotating black hole and a geometrical thin base
of the jet one gets small temperatures of electrons and small Thompson
optical depths which can not explain the soft X-ray slope inferred
from data points. The Compton parameter $y$ would be less than a
general average value of $0.5$ deduced by Mannheim et al. (1995).  The
condition $q_{j}>q_{m}$ comes from the energy and mass conservation
laws along the jet. These physical conditions necessary to describe
the jet base medium will be discussed in detail in a future paper.

\section{Summary }

We have studied a symbiotic jet-disk system stationary in time. The
jet is fed by a geometrically thin disk and we assumed that the inner
part of the "standard" disk contains all energetic conditions
necessary to form and to maintain a stable jet's configuration. The
jet energy has the upper bound corresponding to the gravitational
potential energy lost by the infalling gas between $R_{ms}$ and
$R_{jet}$.

 The structure of the accreting disk is modified by the presence of
the jet.  The mass and angular momentum extraction from the disk makes
it thinner and denser. The disk becomes cooler, especially in the
regions close to the jet.  Between the radius of the marginally stable
circular orbit and the radius where it is assumed that a jet has the
exterior edge $R_{jet}$ we do not have a disk-structure any longer. A
mass rate $q_{m} \dot M$ is pushed out into a powerful jet. We neglect
the influence of the magnetic field in the structure of the disks and
the jets. A very important step remains to be made. We would like to
study in more details the vertical structure of a relativistic disk
with a jet starting at its inner edge. A detailed analysis of a
vertical structure of a relativistic disk surounding a Schwarzschild
black hole has been done by D\"orrer (1995).

 We demonstrate that our disk model driving a jet in the innermost
dense regions very close to a maximal Kerr black hole can replace all
the simple models of the accretion thin disk surrounding a
Schwarzschild black hole.

 The thickness of the jet given by the dimensionless parameter $r_{j}$
is limited by the necessity to have sufficient UV photon flux in order
to explain the high luminosity of the disk and the unsaturated
componized photons in the base of jet.

 The presence of the jet cuts off the high energy part of the spectrum
of the disk without any jet (in the soft X-ray range) and its hot base
comptonizes the photons coming from the disk. The total energy carried
out by the jet is strongly dependent on the mass and angular momentum
of black hole.

   We have shown that there is a strong correlation between the disk
and the jet guided by the conservation laws of the mass, angular
momentum and energy. The jet-disk connection leads to interesting
models which explain the spectra of the quasars over the whole range
of frequency. The assumption that the disk accretion is the machine
which provides the energy in AGNs is the basic ingredient for all
models mentioned.

\begin{acknowledgements} 
A.D. would like to acknowledge receipt of a Max Planck fellowship
which makes her work possible. We wish to thank Drs. H. Falcke,
K. Mannheim and J\"org Rachen for intense and helpful
discussions. P.B. would like to thank Drs. M. Romanova and R. Staubert
for long and intense discussions of these points.

\end{acknowledgements}

\end{document}